# Wavelength independent detection and discrimination of a laser with a sensitivity of 140fW using self-coherence

**David M. Benton, Andrew Ellis and Yiming Li.**

*Aston Institute of Photonic Technologies, Aston University, Birmingham, UK. B4 7ET*
*d.benton@aston.ac.uk

## Abstract

Wavelength independent detection and discrimination of laser radiation has been performed by detecting the 'self coherence' of an incoming light source using interferometry with photon sensitive detectors. The system successfully discriminates between coherent and incoherent sources. Detection of scattered laser light has been performed across the visible spectrum with no filters, internal or external source comparisons. The ultimate detection sensitivity was shown to be 140fW for a continuous wave (CW) HeNe laser at 632nm. We believe this to be the most sensitive wavelength independent detection of a CW laser so far reported.

## 1. Introduction

Lasers are so thermodynamically improbable that only unusual astronomical scale circumstances allow lasing to be produced in nature[1]. Lasers are a technological creation and thus the presence of laser radiation is an indicator of technology usage. It is for this reason that the search for extraterrestrial intelligence (SETI) considers laser radiation to be a 'technosignature' [2] and is actively searching for signs of laser emission [3, 4]. In the military world laser radiation must be identified as it could be a targeting laser and a precursor to incoming ordnance [5] or more recently the use of laser weapons. Thus, laser warning receivers (LWR) were developed in the 1980's [6] to enable countermeasure deployment. A significant issue is that the laser beam must strike the detector in order for it to be seen. This is also a limiting factor with optical alignment for systems such as free space optical communications where search patterns are implemented for both source and detector field of view until mutual alignment is achieved [7,8]. This is a limiting step which delays data transfer.

The predominant detection method is to use laser characteristics of brightness in spatial (well defined beams) or spectral (narrow wavelength range) [9-11] regimes. Thus, looking for the presence of a laser beam is done using an imaging system and looking for a bright spot, or using a spectrometer and looking for a bright wavelength. Both approaches become more difficult and expensive at non-visible wavelengths, and both approaches are limited by the level of background illumination. However, a bright source is not necessarily a laser source and using the laser property of coherence has proved both more sensitive than brightness detection and capable of discrimination [12–15] from bright incoherent sources. The task of a generic laser detection system is to detect any source without prior knowledge of its wavelength or other properties, with no control of the source and no control of the background [10]. Coherent detection methods with high sensitivity such as homodyne or heterodyne detection require *a priori* knowledge and control of the source under detection and are therefore not appropriate in the general case. Interferometric detection methods, however, are viable and in this work we extend the usage of modulated interferometry as a method of detection of source coherence [14,15]. *Coherence* detection is semantically distinct from *coherent* detection which requires a local oscillator!

Detection of pulsed lasers is intrinsically easier than for continuous wave (CW) lasers because of the temporal brightness which allows a further level of discrimination against background light. Sensitivities of such systems tend to be classified but reports of 10-13J [16] for pulse detection sensitivity incorporating a filter and 2.4 nW of power with a periodically pulsed beam [17] using a neuromorphic camera are openly available. Detection of CW lasers has been achieved at a sensitivity of 1nW [14].

Through the use of photon sensitive detectors [18] we demonstrate discrimination and detection of a CW laser at intensity levels well below 1pW by looking for the "self coherence" of the source, arising from the self-coherence function[19]. We use the phrase self-coherence to ensure distinction from techniques that compare a received signal with a reference version such as a local oscillator. This approach is 1000 times more sensitive than previously reported wavelength independent detection of a CW laser. The detection of coherence requires detection of temporal interference fringes, requiring extended detection times. Thus, the minimum detectable coherent intensity is much greater than the absolute intensity sensitivity. This has real-world value in the ability to detect the source indirectly such as through atmospheric scattering or weak reflections thus enabling pre-emptive detection, faster response or alignment correction.

## 2. Materials and Methods

Interferometers such as the Michelson Interferometer divide incoming light into two paths before recombining them to observe interference. If the path length difference between the two paths exceeds the coherence length of the optical source then no interference will be seen - known as a laser unequal path interferometer (LUPI). For broad spectrum lights such as daylight only a few microns of path length difference is required to extinguish interference. Conversely, sources with a coherence length longer than the path asymmetry will display interference and thus can be exposed as coherent.  In addition to path length asymmetry, if one of the path lengths is modulated at a known frequency, such as using a piezo driven mirror, a correlated intensity modulation at the output implies the detection of a laser source. A system for implementing a photon sensitive detector of coherence is shown in Figure 1.

Within the interferometer a piezo mounted mirror is driven by a sinusoidal voltage from an analog output port of a data acquisition system, with amplitude and frequency controlled by a computer.

A multimode optical circulator collects light that is retro-reflected from the interferometer and delivers it to a silicon photomultiplier (SiPM, Thorlabs PDA45), with a second SiPM collecting light exiting through the other interferometer output. Light from the fibres is refocused onto the SiPM detectors to ensure efficient geometric collection.  SiPM output is amplified and sent to counters within the data acquisition unit. The counters are sampled at a regular rate (typically 10kHz) and a time series of counts read by computer (Figure 1b). The channels are then used like a balanced detector and subtracted from each other with the result being Fourier transformed and squared to observe the electrical power at the applied modulation frequency.

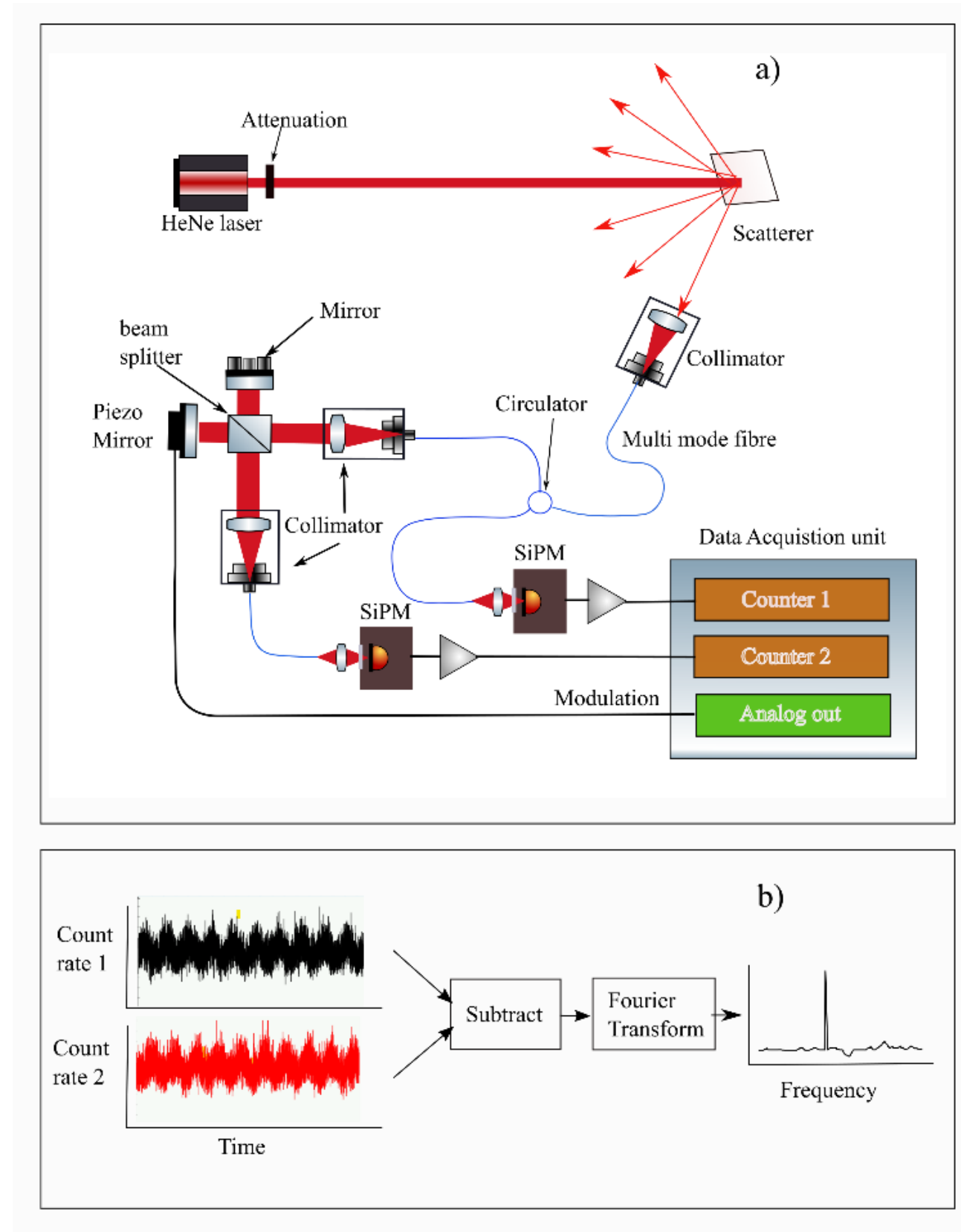

*Figure 1. A schematic diagram of the setup for the photon sensitive detection of coherence. a) Scattered laser light is captured by a collimator and directed via an optical circulator into an interferometer. A piezo mirror is modulated from an analog output signal in a data acquistion unit controlled by a compter (not shown) . Two SiPM detectors in a balanced detector arrangement feed their output pulses to counters in the data acquisition unit. b) Count rates from the two detectors are subtracted before being Fourier transformed to produce a modulation frequency spectrum.*

The signal in the output of a Michelson interferometer with an equal split ratio at the beamsplitters is [20]:

$$I_{out} = \frac{I_0}{4}\varepsilon\left\{1 + Re[\gamma(\tau)]\cos\left(\frac{2\pi}{\lambda}\Delta L\right)\right\} \qquad (1)$$

Where $I_0$ is the input intensity, ε represents losses, $\lambda$ is the wavelength and $\Delta L$ is the path length difference between the 2 interferometer paths. The parameter $\gamma(\tau)$ represents the complex degree of temporal coherence of the source.

$$\gamma(\tau) = \exp\left[-\left(\frac{\pi\Delta f\tau}{2\sqrt{ln2}}\right)^2\right]\exp\left(-2\pi i f_0\tau\right) \qquad (2)$$

Where $f_0$ is the optical frequency, $\Delta f$ is the spectral bandwidth in frequency terms and $\tau$ is the relative time delay of the two arms. In practice the split ratio of beamsplitters is wavelength dependent which results in imperfect cancellation of common intensity noise [21] but this would never be perfect due to the path length difference within the interferometer.

The coherence length of a Gaussian profile source is [22]

$$l_c = \sqrt{\frac{2ln2}{\pi}}\frac{\lambda^2}{\Delta\lambda} \quad (3)$$

Where $\Delta\lambda$ is the spectral bandwidth in wavelength terms.

The path length difference is time-varying as a result of a modulating voltage applied to the piezo mirror in one arm of the Michelson:

$$\Delta L(t) = 2(L_1 - L_2 v\,\rho \sin(2\pi f_m t)) \quad (4)$$

Where $L_1$ and $L_2$ are the distance to each of the mirrors from the centre of the beam splitter, $v$ is the amplitude of the applied modulation voltage of frequency $f_m$ and $\rho$ is the response of the piezo in μm/V. The average path length difference relates to the relative time delay through $\overline{\Delta L} = c\tau$.

Thus, for the two detectors the signal amplitudes vary in antiphase as:

$$S_{1\,or\,2}(t) = \frac{I_0}{4}\varepsilon Q(\lambda)\left\{1 \mp F.\left(e^{-\frac{\overline{\Delta L}}{l_c}}\right)^2 \cos\left(\frac{2\pi}{\lambda}\Delta L(t)\right)\right\} + n_{1\,or\,2} \quad (5)$$

Where $Q(\lambda)$ is the quantum efficiency of the detector, and $n_{1\,or\,2}$ is the background detection rate including dark counts and background photons. The losses $\varepsilon$ include all losses between input and detector. The factor $F$ accounts for the quality of alignment which, when not perfect due to drift etc. causes a reduction in the fringe visibility [21].

The temporal variation of the signal is due only to the modulation of the mirror. For small modulations of the piezo mirror $\Delta L < \lambda/4$ the output oscillates at the modulation frequency for all visible wavelengths. At larger modulation higher harmonics of the modulation frequency can also be observed (see [14] for a more in-depth derivation of the harmonic content of interferometer output signals).

The coherent signal is observed in the modulation spectral frequency domain using the signal from two detectors in a balanced arrangement. The temporal signals from the two detectors are subtracted and then Fourier transformed to provide the power spectrum.

We define our signal to noise ratio (SNR) in the frequency spectrum, as the power at the modulation frequency divided by the mean power at neighboring frequencies. The mean spectral power density is given by the variance of the temporal signal, and the spectral power at the modulation frequency increases as the square of the signal amplitude. Therefore, we expect a quadratic relationship between signal power and noise.

The noise arises from the shot noise in the detector due to signal and background photons, dark counts, detector noise and in some cases amplitude noise of the laser source. These noise contributions have a flat distribution across the detection frequency spectrum. The shot noise $P_s$ for a perfect balanced detector system is:

$$P_s = \sqrt{2\frac{hc}{\lambda}P_{opt}\Delta f} \quad (6)$$

Where $h$ is Planks constant, $c$ is the speed of light, $\lambda$ is the incident wavelength, $P_{opt}$ is the incident optical power and $\Delta f$ is the detection bandwidth.

In the rest of this paper we show 1) discrimination of laser vs LED light of similar brightness, 2) for a real-world scenario, scattered laser light (indirect detection) maintains its coherence and can be discriminated, 3) we probe the coherence detection limit for this system.

## 3. Results

*Coherence detection*

The setup as shown in Figure 1 was used in two configurations, collecting scattered light as shown and also with an attenuated laser fed directly into the collecting collimator for sensitivity measurements.

The first property to demonstrate is that the system really does discriminate and detect sources based on coherence length. To demonstrate this, the system was used with two similar input sources fed into the optical circulator - a HeNe laser attenuated to 0.4μW and a red LED of centre wavelength 632nm directed into a fibre and also delivering 0.4μW. Both sources were operated in a CW mode. The interferometer was modulated at a frequency of 600Hz. The number of detected photons for each detector within a sampling period are recorded in the two counters. Multiple consecutive samples give rise to a time varying count rate which is Fourier transformed to observe the modulation frequency in the interferometer output. Figure 2 shows the modulation frequency power spectrum of the interferometer output light for the laser and LED sources with a sampling rate of 4kHz. The detection bandwidth for the SiPMs is 10MHz thus the shot noise power from equation (6) is 1.6nW. The path length asymmetry also makes it impossible to get perfect common mode noise cancellation [21]. We can clearly see that the coherent source produces a modulating signal at the modulation frequency of 600Hz, whereas the incoherent LED does not. The HeNe laser has a coherence length of around 2mm whereas the LED has a coherence length of 25µm. The actual path length difference within the interferometer is unknown, is estimated to be around 200µm, but is clearly large enough to suppress any modulation arising from incoherent sources.

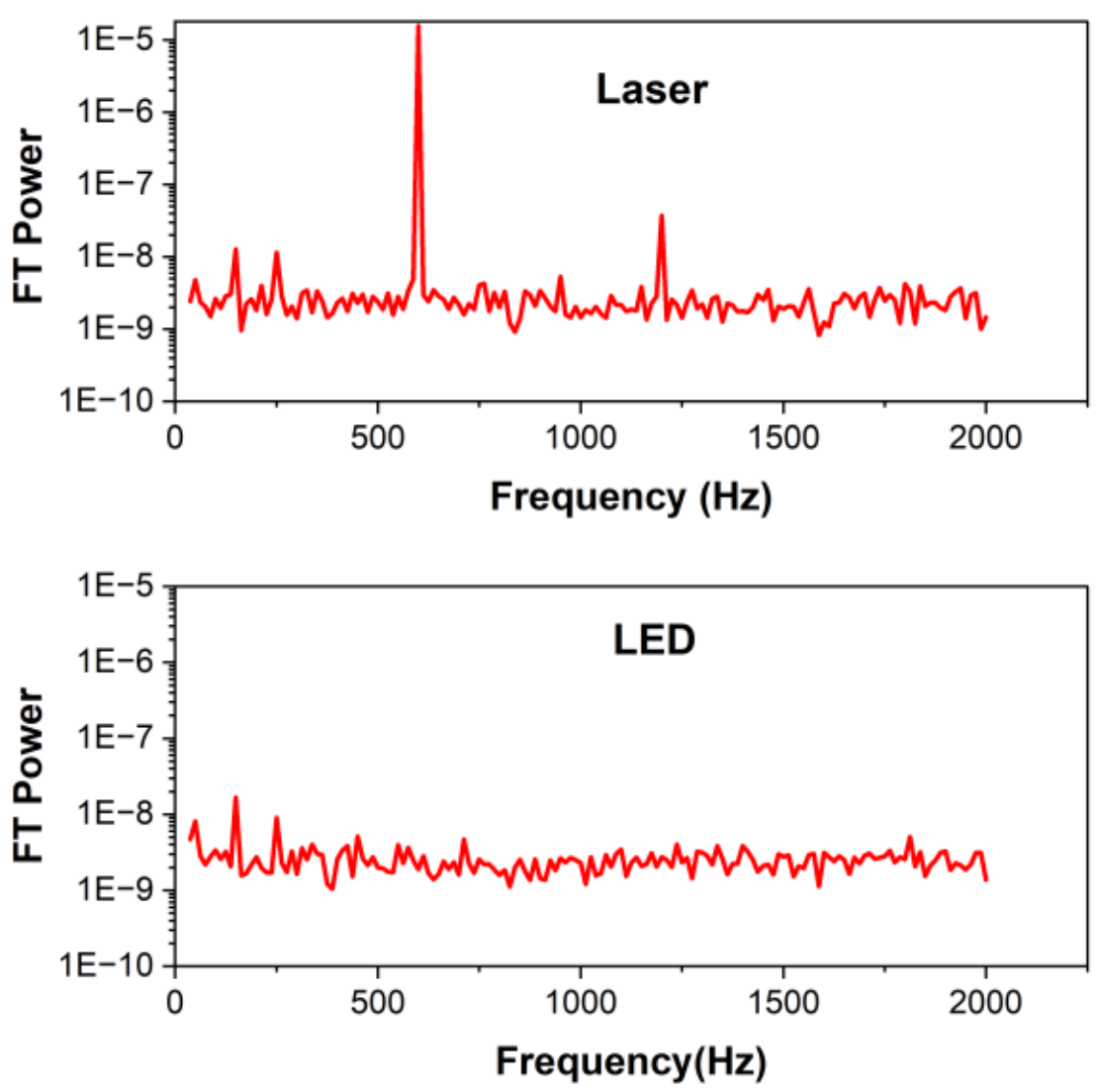

*Figure 2.Fourier Transform (FT) power spectra for a laser and LED with similar wavelengths and similar optical power with a modulation frequency of 600Hz applied to the interferometer piezo mirror.*

### *Wavelength independence*

A generic detector of coherence must be able to detect coherent sources with independence from the source and therefore no filters or wavelength dependency (such as a local oscillator for heterodyne detection). It must also operate with no influence over the environment or operational circumstances. Thus unpredictable incoherent backgrounds are the norm and the most likely route into the detector system in real world usage is via scattering.

Figure 3 shows the detected signals originating from different wavelength lasers scattered from a paper target and collected by a 25mm diameter optical head focusing into a multimode fibre (100µm core) from a distance of 1.6m. This gives a geometrical loss factor for light collection of 3x10-5. Incident lasers were attenuated down to powers of 4µW resulting in around 120pW being collected and directed into the interferometer system. The interferometer was modulated at a frequency of 900Hz. These plots were chosen to highlight some interesting aspects of operation. Firstly, no changes were made to the system between wavelengths, highlighting the truly wavelength independent nature of the coherence detection system. These measurements were taken in a lit laboratory at a sampling rate of 10kHz and show that temporal coherence is maintained through diffusive scattering. The red plot shows the detected signal at a modulation frequency of 900Hz from a scattered HeNe laser but additionally shows a peak at around 3.4kHz which originates from an LED lamp. This is not a coherent signal but an amplitude modulation of the lamp. Such modulations can be observed, particularly when the 2 arms of the interferometer do not have equal efficiency and balanced detector subtraction is not complete. It shows the importance of selecting an easily identifiable modulation frequency. The green plot is the detection of a green laser at 532nm which carries a lot of intensity noise and has a coherence length of 370µm. It shows the frequency components of main room lights at 50Hz and again is easily distinguished. The blue plot is detection of a blue laser at 405nm with a coherence length of 217 µm. In this case the signal is much weaker because scattering paper absorbs blue light and fluoresces brightly reducing signal and increasing background photons. The shot noise power for the HeNe in this case is 27pW and dominated by the laser power. Differences in beamsplitter reflectivity and polarization sensitivity mean that the noise cancelling effect of the balanced arrangement is again imperfect [21], but still beneficial. These measurements are presented to firstly show that the system is truly wavelength independent but, secondly can be practically utilised, not just in a well-controlled lab environment.

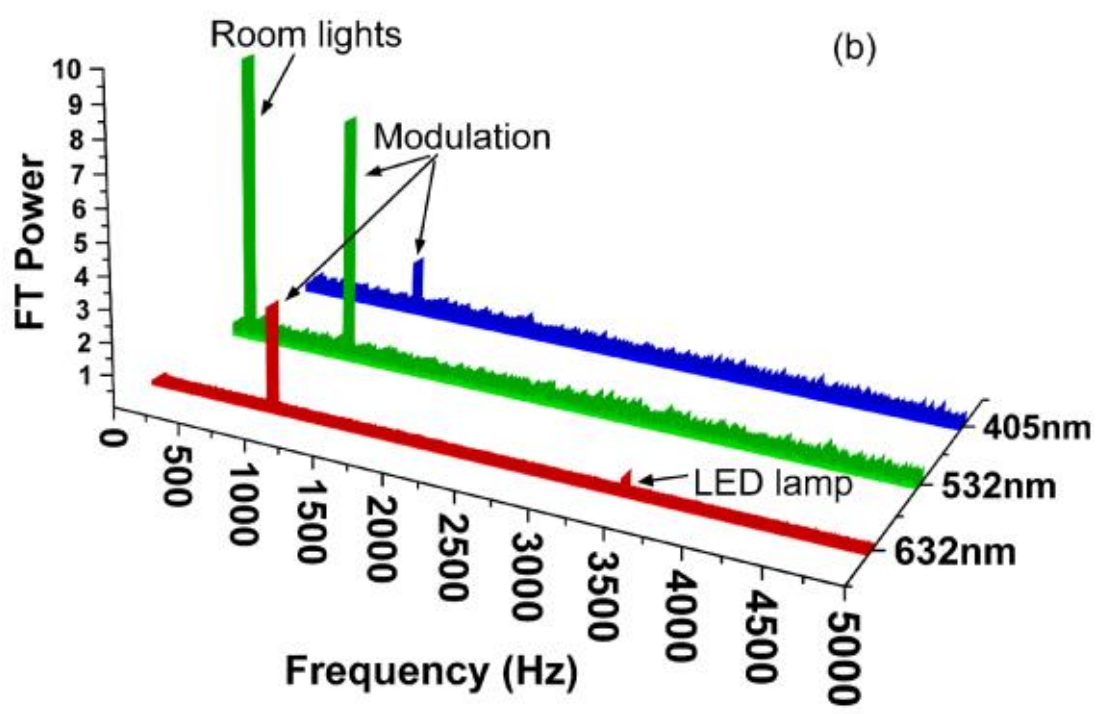

*Figure 3. Fourier transforms of the detected temporal signals from 3 laser wavelengths with different background lighting conditions.*

*Sensitivity*

To assess the sensitivity of the system a HeNe laser was focused directly into the multi-mode optical circulator via a collimator. The laser source was attenuated using calibrated neutral density filters. Sampling at a rate of 10kHz, collecting a batch of 10000 samples and integrating for 4 batches has an integration time of 4s. Coherence detections were performed with input laser powers from 0.07 to 4.5 pW. We were able to observe sub-pW laser detection at 632nm with a SNR of 2.8±0.3 for a laser intensity of 0.14pW. Photon rates in each detector were 25kcps and were the dominant noise source.

The system was modelled according to equation 5. The calculated signal values and noise levels were converted from intensity to expected photon number and used to randomly select a discrete photon number from a Poisson distribution with a mean equal to the photon rate per sample. The observed count rates in comparison to the expected photon rates suggest the system efficiency is less than 2% which arises from a quantum efficiency of Q(λ)=22% and losses from 15 uncoated glass surfaces and an oversize beam entering the interferometer ($\boldsymbol{\varepsilon} = \mathbf{0.1}$). A contrast factor of F=15% produced equivalent SNR values that align with the measurements. A series of expected SNR values were calculated for increasing laser input powers. Measured SNR values along with modelled values of SNR are shown in Figure 4. For consistent, reliable performance reference level we choose a system sensitivity at an SNR value of 5 which corresponds to detection at 0.25pW.

Using this model with improved but realistic values for surface reflection coefficients (1%), reduced dark counts (5k cps) but the same background level and improved contrast (75%) we can estimate that a future version of this system should be able to detect laser powers of 8fW (SNR=5), with performance in the current laser power regime shown in as the line labelled ‘Potential’.

This work is grounded in the requirement to detect a non-specific laser wavelength with high sensitivity. We have concentrated on CW lasers as these are most difficult to detect without the benefit of temporal brightness that is present for pulsed lasers. Practicality is an important issue in this field. Whilst laser specific schemes such as filters or heterodyne detection could offer better levels of sensitivity, they require a priori knowledge of the lasers to be detected, which becomes impractical as the number of potential lasers continues to increase.

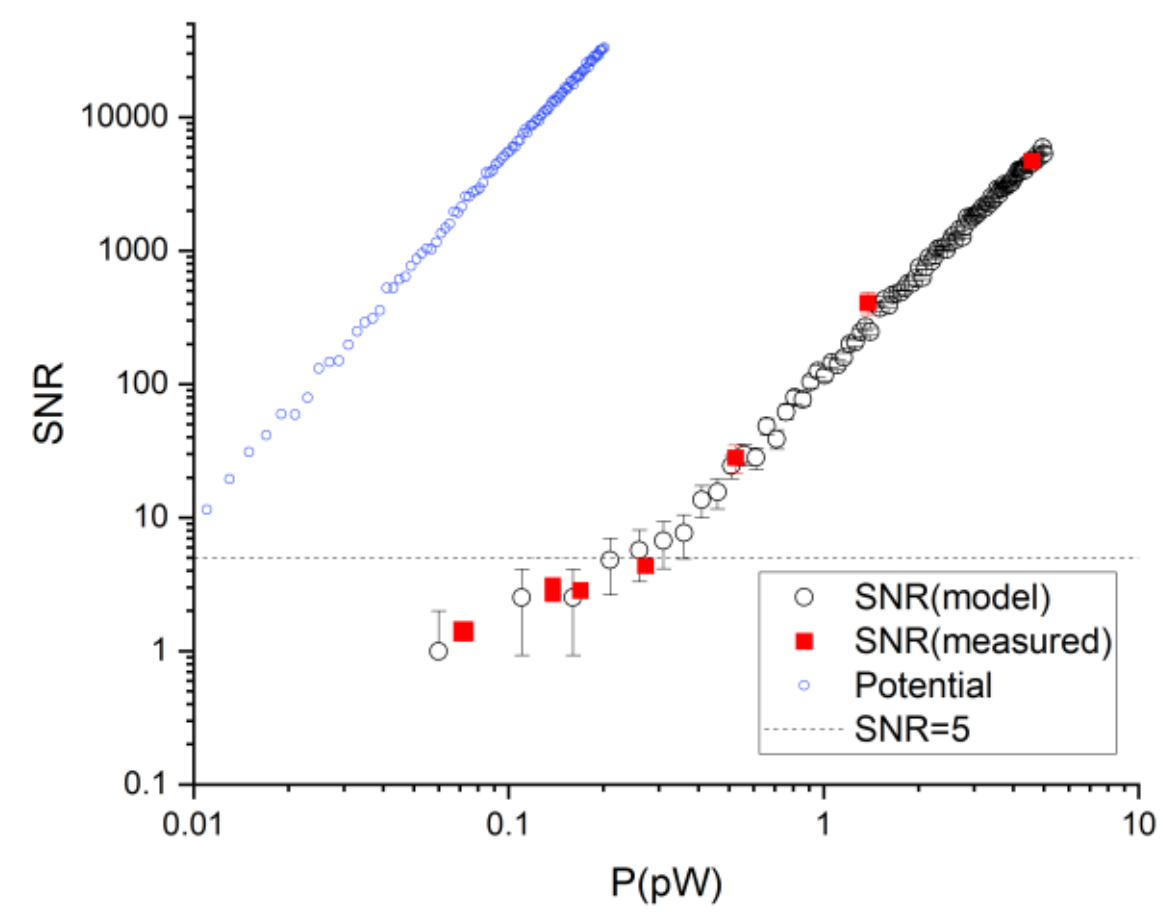

*Figure 4. Modelled and measured SNR values vs laser power (P) with a modelled potential performance. The dotted line represents SNR=5.*

## 4. Discussion

A single system with wide spectral applicability is an attractive offering. The theme of practicality continued with how the system was tested. A total collection time of 4 seconds was used because detection of a laser must be done in a timely fashion in order to be useful. This does of course limit the system detection sensitivity. By collecting more samples with longer integration times we could see the sensitivity improve but this is application specific. For example, in searching for laser emission emanating from the stars it would be appropriate to integrate for hours potentially improving sensitivity a hundred-fold, but this is not useful in observing, for example, the irradiation of civil aircraft which requires timescales of a few seconds at most.

This is not to say that this technique would not benefit from some specificity. The biggest issue affecting sensitivity is the background level and any amount of spectral filtering will reduce background and increase SNR. This could be of use in aligning free space optical systems where a limited number of wavelengths will be used but it is not appropriate to fix on any one wavelength. The ability to detect scattered or off axis laser radiation could be helpful in speeding up the acquisition of alignment by detecting the incoming laser more quickly, enabling more data transfer in a limited time window, or more stable connections in difficult environmental conditions.

Whilst the lasers used here have been visible, the technique is generic and could equally apply to infrared or ultraviolet wavelengths, with the understanding that current detector technology limits the photon sensitivity in these spectral regions. It is also worth commenting that partially coherent sources can produce modulating output if their coherence length is less than the interferometry path length asymmetry e.g, atomic emission lamps can have narrow lines so some care should be taken before assigning a modulating output to a laser input - this can be easily rectified by examining the wavelength. Pulsed lasers will show a component at the pulse repetition frequency if that is less than sampling Nyquist frequency. Higher frequency pulse rates can contribute to output at the modulation frequency. Such effects have been observed but the focus in the work has been on continuous lasers.

## 5. Conclusions

In conclusion we have demonstrated wavelength independent detection and discrimination of laser radiation based upon the observation of source coherence rather than source brightness. This system involves no filters, no local oscillators (as per homodyne or heterodyne detection) and no significant physical movements such as for Fourier transform spectroscopy. Using silicon photomultipliers we have demonstrated sensitive detection of scattered laser radiation right across the visible spectrum for weak continuous wave lasers. In testing for the system sensitivity we have detected a HeNe laser at an intensity level of 140fW with a signal to noise ratio of 2.5(±0.3) and we believe this to be the most sensitive generic detection of CW lasers ever reported. This was performed using general equipment found within a lab. Modelling suggests that using tailored components could result in an ultimate sensitivity for this system of 8fW.